\documentclass[conference]{IEEEtran}
\IEEEoverridecommandlockouts

\usepackage{cite}
\usepackage{amsmath,amssymb,amsfonts}
\usepackage{algorithmic}
\usepackage{algorithm}
\usepackage{graphicx}
\usepackage{textcomp}
\usepackage{xcolor}
\usepackage{booktabs}
\usepackage{multirow}
\usepackage{tikz}
\usetikzlibrary{shapes,positioning,arrows.meta}

\def\BibTeX{{\rm B\kern-.05em{\sc i\kern-.025em b}\kern-.08em
    T\kern-.1667em\lower.7ex\hbox{E}\kern-.125emX}}

\begin{document}

\title{GPU-Accelerated Real-Time Software Defined Radio-Based Orthogonal Time Frequency Space Network-Coded Cooperation System: Hardware Implementation
}

\author{
  \IEEEauthorblockN{You-Yu Huang}
  \IEEEauthorblockA{\textit{Dept. of Communication Engineering} \\
  \textit{National Chung Cheng University}\\
  Chiayi, Taiwan \\
  jack1003@alum.ccu.edu.tw}
  \and
  \IEEEauthorblockN{Yu-Ming Yeh}
  \IEEEauthorblockA{\textit{Dept. of Communication Engineering} \\
  \textit{National Chung Cheng University}\\
  Chiayi, Taiwan \\
  yklibo0226@gmail.com}
  \and
  \IEEEauthorblockN{Jen-Yi Pan\IEEEauthorrefmark{1}}
  \IEEEauthorblockA{\textit{Dept. of Communication Engineering} \\
  \textit{National Chung Cheng University}\\
  Chiayi, Taiwan \\
  jypan@ccu.edu.tw}
  \thanks{\IEEEauthorrefmark{1}Corresponding author}
}

\maketitle

\begin{abstract}
While Orthogonal Time Frequency Space (OTFS) modulation offers robust reliability for 6G vehicular networks, standalone links suffer from blockages, and existing Software-Defined Radio (SDR) testbeds are bottlenecked by complex Delay-Doppler (DD) equalizers. This paper presents a real-time Decode-and-Forward Network-Coded Cooperation OTFS (OTFS-NCC) prototype implemented on consumer hosts and USRP B210 SDRs. Operating over a five-node TDD (Time-Division Duplexing) topology, our framework improves spectral efficiency by 33\% over conventional relaying while mitigating error propagation via an enhanced Gaussian Approximate Message Passing Algorithm (GA-MPA). To support a 2 MHz baseband rate, we devise a hardware-algorithm decoupled GPU (Graphics Processing Unit) architecture using 1D memory mapping and transcendental function clipping, compressing the simulated Real-Time Factor (RTF) from 4.37 to 0.89. RF-conducted Hardware-in-the-Loop validation under 3GPP EVA70 (Extended Vehicular A model) fading confirms sustained zero-packet-drop real-time demodulation over 60-second test runs across a large 112-by-64 DD grid.
\end{abstract}


\section{Introduction}
In high-speed 6G vehicular-to-everything (V2X) scenarios, conventional multi-carrier waveforms suffer from severe inter-carrier interference \cite{6G_Intro, V2X_Intro}. To overcome this, Orthogonal Time Frequency Space (OTFS) modulation emerges as a transformative paradigm \cite{OTFS_Intro}, multiplexing symbols in the Delay-Doppler (DD) domain to compact doubly dispersive channels into a sparse, 2D grid representation.

Despite its waveform superiority, a standalone direct OTFS link remains intrinsically vulnerable to severe physical blockages and deep shadowing. While Cooperative Communication (CC) effectively exploits spatial diversity via intermediate relays to bypass obstructions \cite{CC_Intro}, conventional half-duplex relaying incurs a severe $50\%$ penalty in spectral efficiency. Integrating Network Coding (NC) with cooperative relaying—termed Network-Coded Cooperation (NCC)—resolves this trade-off by allowing intermediate relays to algebraically superimpose packets from distributed sources \cite{NCC_Intro}. Consequently, synthesizing an OTFS-based NCC framework (OTFS-NCC) stands out as an optimal architectural candidate for 6G doubly dispersive scenarios \cite{OTFS_outage_ana, OTFS_NCMA}.

However, transitioning OTFS cooperative frameworks from mathematical theory to live Radio Frequency (RF) hardware testbeds introduces severe engineering hurdles. Existing state-of-the-art SDR implementations by Rahman \textit{et al.} \cite{Rahman_2023_OTFS_SDR} and Rasendria \textit{et al.} \cite{Rasendria_2025_OTFS_SDR} successfully demonstrated over-the-air OTFS transmissions; yet, both prototypes were forced to restrict their baseband configurations to micro-grids of $M = 64$ subcarriers and $N = 16$ time slots ($1,024$ Resource Elements) at a narrow $1\text{ MHz}$ bandwidth. In practical SDR transceivers, general-purpose host processors are heavily bottlenecked by standard Message Passing Algorithm (MPA) equalizers, whose intractable computational complexity, massive memory footprint, and frequent transcendental function evaluations ($\tanh$) yield prohibitive Real-Time Factors ($\text{RTF} \gg 1$), resulting in baseband buffer overflows and catastrophic packet losses.

To bridge the gap between theoretical 6G waveform design and empirical realization, this paper designs, optimizes, and validates an end-to-end real-time OTFS-NCC hardware prototype.

\section{System Model and Proposed DF-NCC Protocol}

\subsection{OTFS DD-Domain Grid and Fading Model}
\begin{figure}[htbp]
    \centering
    \includegraphics[width=0.88\linewidth]{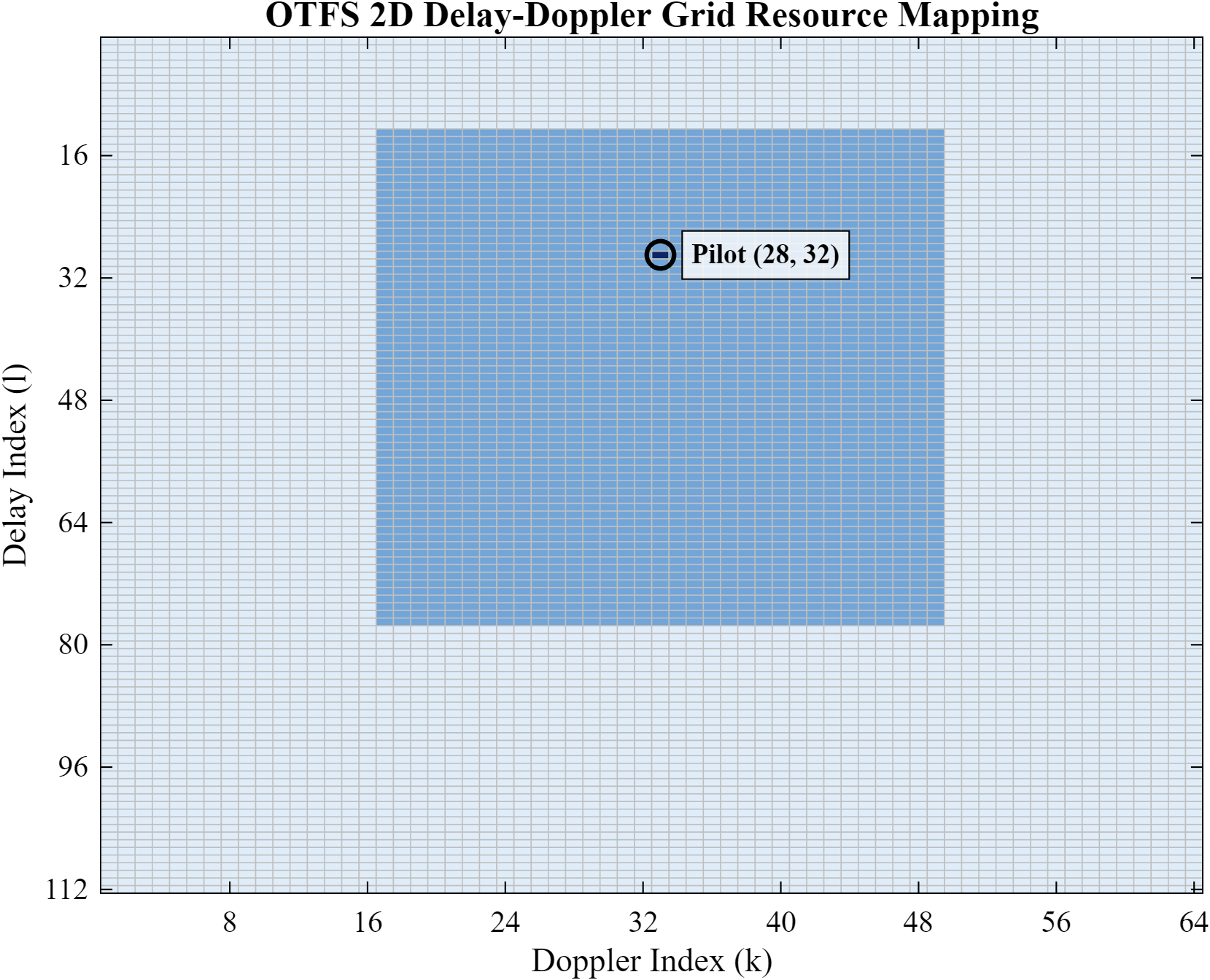}
    \caption{Delay-Doppler (DD) domain grid representation and frame resource allocation structure of the OTFS waveform.}
    \label{fig:otfs_grid}
\end{figure}

Consider an OTFS frame spanning a 2D DD grid of size $M \times N$, illustrated in Fig. \ref{fig:otfs_grid}, where $M$ is the number of subcarriers with spacing $\Delta f$, and $N$ is the number of time slots with duration $T = 1/\Delta f$. Information bits are mapped to BPSK symbols $x[l,k]$ at delay index $l \in [0, M-1]$ and Doppler index $k \in [0, N-1]$. 

The DD symbols are transformed to the Time-Frequency (TF) domain via the Inverse Symplectic Finite Fourier Transform (ISFFT). Subsequently, the Heisenberg transform applies rectangular multi-carrier modulation to generate the time-domain waveform $s(t)$. After traversing a doubly dispersive vehicular channel with $P$ resolvable multipaths, explicitly considering fractional Doppler leakage, the effective baseband received signal $y[l,k]$ is modeled as \cite{OTFS_non_ideal}:
\begin{equation}
    y[l, k] = \sum_{i=1}^{P} h_i e^{j\phi_i(l, k)} \mathcal{G}_i(l, k) x\left[ [l - l_i]_M, [k - k_i]_N \right] + w[l, k],
\end{equation} 
where $h_i$ is the complex path gain, $(l_i, k_i)$ denote the delay and Doppler index offsets of the $i$-th path, $\mathcal{G}_i(l, k)$ represents the fractional Doppler spatial leakage function, and $w[l, k]$ is additive white Gaussian noise (AWGN).

\subsection{Five-Node Three-Slot TDD Topology}
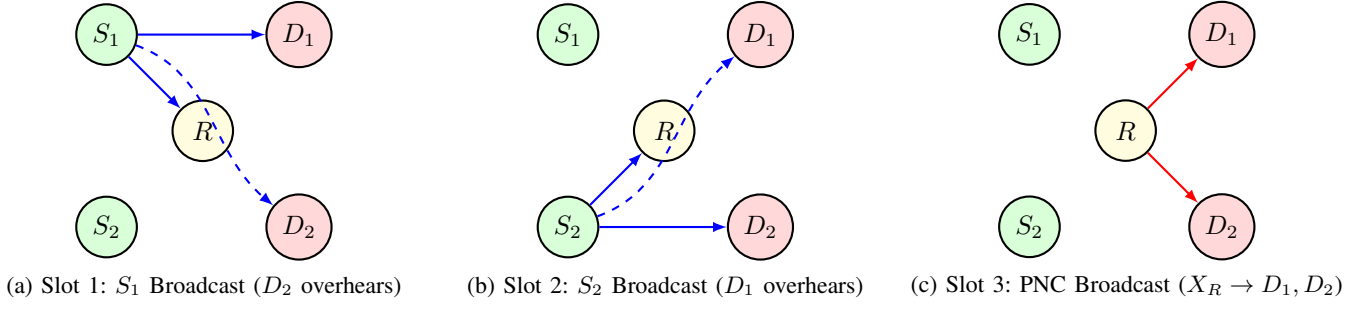
\begin{figure*}[htbp]
    \centering
    \begin{minipage}{0.32\textwidth}
        \centering
        \begin{tikzpicture}[scale=0.6, node distance=18mm, main/.style={draw, circle, thick, minimum size=8mm}]
            \node[main, fill=green!15] (s1) {$S_1$};
            \node[main, fill=yellow!15] (r) [below right of=s1] {$R$};
            \node[main, fill=green!15] (s2) [below left of=r] {$S_2$};
            \node[main, fill=red!15] (d1) [above right of=r] {$D_1$};
            \node[main, fill=red!15] (d2) [below right of=r] {$D_2$};
            \draw[-latex, thick, blue] (s1) -- (r);
            \draw[-latex, thick, blue] (s1) -- (d1);
            \draw[-latex, thick, blue, dashed] (s1) to[out=-20, in=140] (d2);
        \end{tikzpicture}
        \centerline{\small (a) Slot 1: $S_1$ Broadcast ($D_2$ overhears)}
    \end{minipage}\hfill
    \begin{minipage}{0.32\textwidth}
        \centering
        \begin{tikzpicture}[scale=0.6, node distance=18mm, main/.style={draw, circle, thick, minimum size=8mm}]
            \node[main, fill=green!15] (s1) {$S_1$};
            \node[main, fill=yellow!15] (r) [below right of=s1] {$R$};
            \node[main, fill=green!15] (s2) [below left of=r] {$S_2$};
            \node[main, fill=red!15] (d1) [above right of=r] {$D_1$};
            \node[main, fill=red!15] (d2) [below right of=r] {$D_2$};
            \draw[-latex, thick, blue] (s2) -- (r);
            \draw[-latex, thick, blue] (s2) -- (d2);
            \draw[-latex, thick, blue, dashed] (s2) to[out=20, in=-140] (d1);
        \end{tikzpicture}
        \centerline{\small (b) Slot 2: $S_2$ Broadcast ($D_1$ overhears)}
    \end{minipage}\hfill
    \begin{minipage}{0.32\textwidth}
        \centering
        \begin{tikzpicture}[scale=0.6, node distance=18mm, main/.style={draw, circle, thick, minimum size=8mm}]
            \node[main, fill=green!15] (s1) {$S_1$};
            \node[main, fill=yellow!15] (r) [below right of=s1] {$R$};
            \node[main, fill=green!15] (s2) [below left of=r] {$S_2$};
            \node[main, fill=red!15] (d1) [above right of=r] {$D_1$};
            \node[main, fill=red!15] (d2) [below right of=r] {$D_2$};
            \draw[-latex, thick, red] (r) -- (d1);
            \draw[-latex, thick, red] (r) -- (d2);
        \end{tikzpicture}
        \centerline{\small (c) Slot 3: PNC Broadcast ($X_R \to D_1, D_2$)}
    \end{minipage}
    \vspace{1mm}
    \caption{Proposed three-slot TDD cooperative transmission protocol over the five-node wireless topology.}
    \label{fig:tdd_topology}
\end{figure*}

Conventional cooperative relaying requires four time slots to serve two users, forfeiting 50\% spectral efficiency. Our proposed OTFS-NCC protocol overcomes this via a five-node topology operating over three Time-Division Duplexing (TDD) slots, as depicted in Fig. \ref{fig:tdd_topology}.

During Slot 1 and 2, $S_1$ and $S_2$ broadcast sequentially. While $R$ and $D_1$ receive $S_1$'s frame, $D_2$ silently overhears due to the broadcast medium. Symmetrically, $D_1$ overhears $S_2$ during Slot 2. In Slot 3, relay $R$ decodes both incoming frames. Upon reconstructing clean BPSK grids $\hat{X}_1$ and $\hat{X}_2$, $R$ performs Physical-layer Network Coding (PNC) via algebraic superposition \cite{NCC_PNC}:
\begin{equation}
    X_R[l, k] = \frac{1}{2}\hat{X}_1[l, k] + \frac{1}{2}\hat{X}_2[l, k],
\end{equation}
scaling by $0.5$ to prevent RF power amplifier (PA) saturation clipping, and broadcasts $X_R$ simultaneously to $D_1$ and $D_2$.

\subsection{Physical-Layer SIC and Shadow Interference Cancellation}
Consider destination $D_1$ targeting $S_1$'s data. In Slot 3, the received broadcast frame $Y_{R \to D_1}^{(3)}$ contains the desired $S_1$ signal heavily corrupted by superimposed $S_2$ interference. However, utilizing the priori frame overheard from Slot 2, $D_1$ already possesses a highly reliable decoded BPSK grid $\hat{\mathbf{X}}_2$.

To isolate $S_1$'s fading signal, $D_1$ executes a 2D circular convolution between $\hat{\mathbf{X}}_2$ and the estimated $R \to D_1$ multipath channel parameters to synthesize a "shadow interference grid" $\mathbf{I}_2$. Exploiting the cyclic property of the DD grid, this discrete convolution is efficiently resolved via 2D circular matrix shifts, as outlined in Algorithm \ref{alg:shadow_convolution}.

\begin{algorithm}[htbp]
\caption{2D Delay-Doppler Shadow Channel Convolution}
\label{alg:shadow_convolution}
\begin{algorithmic}[1]
\REQUIRE Prior decoded interferer grid $\hat{\mathbf{X}}_2 \in \mathbb{C}^{M \times N}$, Channel paths $P$, Gain list $\mathbf{h}_{list}$, Delay shifts $\mathbf{l}_{idx}$, Doppler shifts $\mathbf{k}_{idx}$
\ENSURE Shadow interference grid $\mathbf{I}_2 \in \mathbb{C}^{M \times N}$
\STATE $\mathbf{I}_2 \leftarrow \mathbf{0}_{M \times N}$
\FOR{$p = 1$ \TO $P$}
    \IF{$\mathbf{h}_{list}[p] == 0$}
        \STATE \textbf{continue}
    \ENDIF
    \STATE // \textit{Perform 2D circular shift along Doppler and Delay dimensions}
    \STATE $\mathbf{X}_{shifted} \leftarrow \text{CircShift}\left(\hat{\mathbf{X}}_2, \, [\mathbf{k}_{idx}[p], \, \mathbf{l}_{idx}[p]]\right)$
    \STATE // \textit{Accumulate multipath interference weighted by channel gain}
    \STATE $\mathbf{I}_2 \leftarrow \mathbf{I}_2 + \mathbf{h}_{list}[p] \cdot \mathbf{X}_{shifted}$
\ENDFOR
\RETURN $\mathbf{I}_2$
\end{algorithmic}
\end{algorithm}

Once $\mathbf{I}_2$ is synthesized, Successive Interference Cancellation (SIC) is executed via direct algebraic subtraction:
\begin{equation}
    \tilde{Y}_{D_1}^{(3)}[l, k] = Y_{R \to D_1}^{(3)}[l, k] - \frac{1}{2} \mathbf{I}_2[l, k].
\end{equation}
Finally, soft log-likelihood ratios (LLRs) extracted from the SIC-cleaned relay link $\tilde{Y}_{D_1}^{(3)}$ are merged with LLRs from the direct link ($S_1 \to D_1$ received in Slot 1) via Maximum Ratio Combining (MRC).

\section{End-to-End SDR Transceiver Architecture}
\begin{figure}[htbp]
    \centering
    \includegraphics[width=0.92\linewidth]{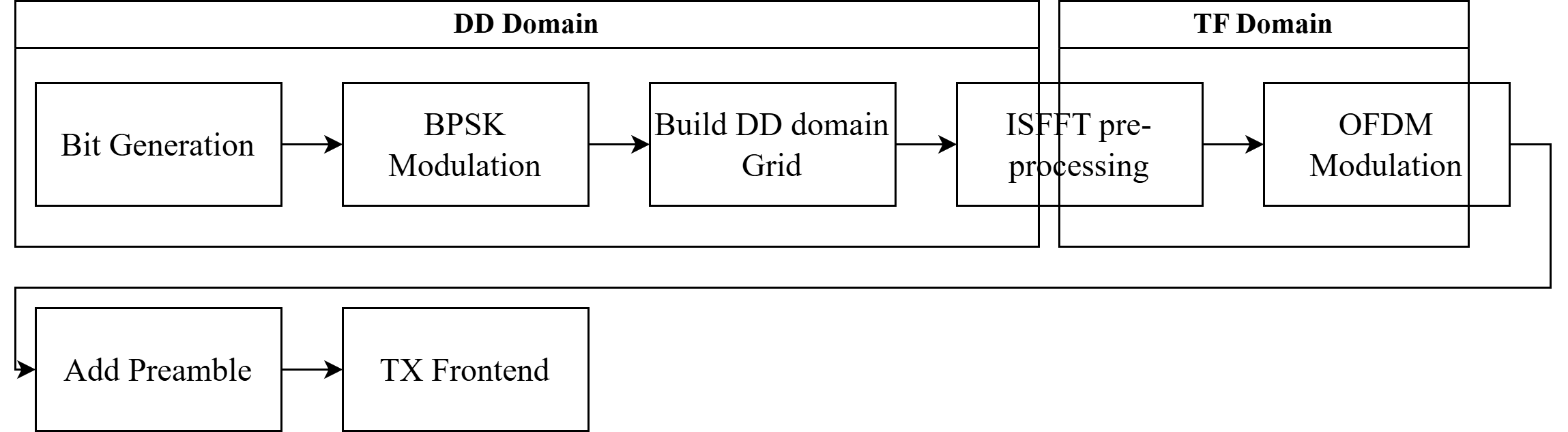}
    \vspace{2mm}
    \centerline{\small (a) SDR Transmitter Baseband Processing Pipeline}
    \vspace{4mm}
    \includegraphics[width=0.92\linewidth]{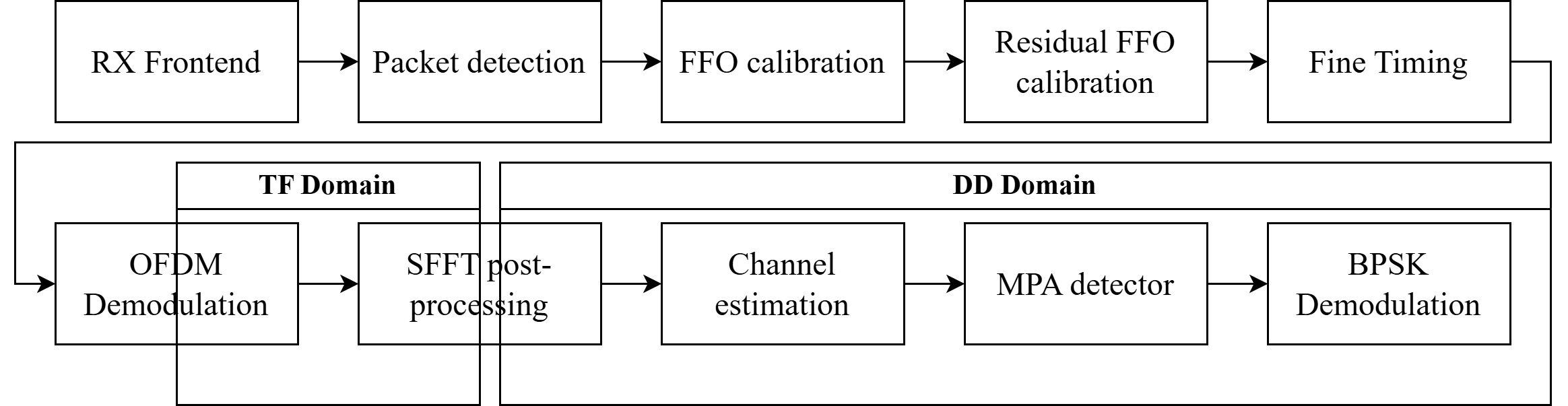}
    \vspace{2mm}
    \centerline{\small (b) SDR Receiver Baseband Processing Pipeline}
    \caption{Host-to-hardware functional block diagrams of the end-to-end OTFS-NCC SDR prototyping testbed.}
    \label{fig:tx_rx_structure}
\end{figure}

To validate the protocol under real RF hardware impairments, we construct a fully functional baseband transceiver deployed on NI USRP B210 SDRs driven by heterogeneous computing hosts. The hardware specifications and software configurations ensuring testbed reproducibility are summarized in Table \ref{tab:hw_sw_spec}. The symmetrical baseband execution pipelines are detailed in Fig. \ref{fig:tx_rx_structure}.

\begin{table}[htbp]
    \centering
    \caption{Experimental Platform Hardware Specs and Software Environment}
    \begin{tabular}{l l}
        \toprule
        \textbf{Category / Item} & \textbf{Specification / Version Info} \\
        \midrule
        Host CPU & Intel Core i7-13700KF (16 Cores / 24 Threads) \\
        Host GPU & NVIDIA GeForce RTX 3070 Ti (8 GB GDDR6X) \\
        System RAM & 64 GB DDR4 3600 MT/s (Dual Channel) \\
        Storage & PCIe Gen4 NVMe M.2 SSD \\
        Operating System & Windows 11 (25H2) \\
        MATLAB / Simulink & R2025b (Simulink Accelerator Mode) \\
        CUDA Engine & CUDA v13.2 \\
        SDR Driver & UHD v4.6.0.0 \\
        SDR Hardware & NI USRP B210 ($f_c = 2.1\text{ GHz}, F_s = 2\text{ MHz}$) \\
        \bottomrule
    \end{tabular}
    \label{tab:hw_sw_spec}
\end{table}

\subsection{HIL Hardware Configuration and SCO Evasion Strategy}
Operating at a hardware master clock of $32\text{ MHz}$ with an interpolation/decimation factor of $16$, the effective baseband sampling rate converges strictly to $F_s = 2\text{ MHz}$ at a $2.1\text{ GHz}$ carrier frequency. 

Crucially, practical distributed wireless testbeds are notoriously plagued by Sampling Clock Offset (SCO) and Carrier Frequency Offset (CFO) driven by independent crystal oscillator tolerances across transceivers. To isolate and rigorously evaluate physical waveform resilience against dynamic vehicular multipath without compounding uncalibrated hardware clock drift, our prototyping framework adopts a closed Hardware-in-the-Loop (HIL) topology. By utilizing the dual RF channels (TX and RX2) of a single USRP B210 transceiver connected via shielded SMA cables and $30\text{ dB}$ attenuators, both transmitter and receiver baseband interfaces share the exact same internal Temperature Compensated Crystal Oscillator (TCXO) reference clock. This architectural design completely eliminates SCO and static hardware frequency offsets from the physical layer, ensuring that empirical error floors strictly reflect doubly dispersive fading challenges.

\subsection{Baseband Grid Configuration and Preamble Design}
As detailed in Table \ref{tab:otfs_parameters}, the OTFS baseband defines an FFT size of $128$ and a Cyclic Prefix length of $64$ samples. Excluding guard bands and DC nulls, the DD grid allocates $M = 112$ subcarriers and $N = 64$ time slots ($7,168\text{ REs}$). A high-power complex pilot pulse $P_{val} = 50(1-1j)/\sqrt{2}$ is injected at anchor $(28, 32)$, surrounded by an asymmetric guard band ($C_{guard} = 2,145\text{ REs}$) to absorb multipath delay spread, yielding an effective payload capacity of $N_{data} = 5,023\text{ REs}$ per frame ($T_{frame} = 6.144\text{ ms}$).

\begin{table}[htbp]
\caption{OTFS Baseband Grid and Multicarrier Parameters}
\label{tab:otfs_parameters}
\centering
\begin{tabular}{lcc}
\hline
\textbf{Parameter Name} & \textbf{Symbol} & \textbf{Value / Configuration} \\
\hline
Physical Sample Rate & $F_s$ & 2 MHz \\
Effective Baseband Rate & $F_{\text{eff}}$ & 1 MS/s ($50\%$ Duty Cycle) \\
FFT Size / CP Length & $N_{\text{fft}}$ / $N_{\text{cp}}$ & 128 / 64 samples \\
Subcarriers / Time Slots & $M$ / $N$ & 112 / 64 (7,168 REs) \\
Pilot Anchor Position & $(l_p, k_p)$ & $(28, 32)$ \\
Complex Pilot Value & $P_{\text{val}}$ & $50(1-1j)/\sqrt{2}$ \\
Guard Band Overhead & $C_{\text{guard}}$ & 2,145 REs \\
Effective Data Payload & $N_{\text{data}}$ & 5,023 REs per frame \\
Frame Duration & $T_{\text{frame}}$ & 6.144 ms \\
Doppler Resolution & $\Delta \nu$ & 162.76 Hz \\
\hline
\end{tabular}
\end{table}

To achieve robust asynchronous frame synchronization over RF thermal noise, the transmitter appends a Chu-sequence preamble structure organized as [STS, LTS-GI, LTS]. The receiver executes a sliding-window Minn autocorrelation timing metric $\mathcal{M}(d)$, whose sharp plateau-free peak guarantees precise frame boundary detection. To quantify continuous RF-conducted frame retrieval without prior packet-boundary knowledge, the receiver implements an \textit{asynchronous double-window buffer strategy}. Specifically, the host interface continuously fetches a sliding baseband observation window spanning twice the frame duration ($2T_{\text{frame}}$). By the cyclic containment principle, an arbitrary incoming timing offset ensures that at least one complete preamble-to-payload sequence is fully captured within the buffer. Once the Minn metric $\mathcal{M}(d)$ identifies the frame boundary, the exact payload is truncated and dispatched to the CUDA equalizer, while the residual buffer samples are flushed. Consequently, while the USRP hardware front-end maintains a continuous physical baseband sampling rate of $F_s = 2\text{ MHz}$, this double-window truncation strategy inherently trades host interface processing throughput for blind frame synchronization robustness. As a result, the effective baseband sample processing rate delivered to the CUDA equalizer converges to $1\text{ MS/s}$ (i.e., a $50\%$ operational duty cycle at the host interface). Coarse fractional CFO is estimated from the phase rotation of Minn correlation peaks ($\Delta\theta = \theta_{CFO}/64$), while fine residual CFO is neutralized via LTS cross-correlation. Finally, fine sample-level timing alignment is resolved via differential LTS cross-correlation incorporating an empirical CP margin back-off ($\Delta_{CP} = -4\text{ samples}$).

\subsection{Real-Time HIL Doubly Dispersive Channel Emulation}
To evaluate transceiver resilience under standardized 6G mobility benchmarks, we embed a Hardware-in-the-Loop (HIL) doubly dispersive channel emulation engine between the SDR interfaces. The emulator synthesizes dynamic multipath fading conforming to the 9-tap 3GPP Extended Vehicular A geometry \cite{3gpp_ts_36_104_v11_2_0} with a maximum Doppler shift of $70\text{ Hz}$ (EVA70). To circumvent host buffer overrun (O) exceptions caused by native double-precision modules, our custom emulator enforces a strict single-precision (FP32) quantization alongside Coder directives (\texttt{\#codegen}) and Blackman Sinc interpolation featuring DC gain normalization. Furthermore, a static $6\text{ dB}$ path-loss degradation is intentionally injected into direct $S \to D$ links.

\section{Hardware-Algorithm Decoupled Acceleration}
While the Gaussian Approximate Message Passing Algorithm (GA-MPA) \cite{GA_MPA_Classic_Paper} linearizes detection complexity from $\mathcal{O}(|\mathbb{A}|^P)$ to $\mathcal{O}(P \cdot |\mathbb{A}|)$, executing belief propagation over loopy graphs on large grids ($112 \times 64$) easily exhausts host memory bandwidth. To achieve real-time baseband demodulation ($\text{RTF} < 1$), we introduce a decoupled hardware-algorithm acceleration architecture.

\begin{algorithm}[htbp]
\caption{Hardware-Optimized 1D Linearized GA-MPA Kernel}
\label{alg:mpa_detector}
\begin{algorithmic}[1]
\REQUIRE Flattened DD observation $\mathbf{y} \in \mathbb{C}^{MN \times 1}$, Channel paths $\{h_p, \Delta l_p, \Delta k_p\}_{p=1}^{P}$, Noise $\sigma^2_w$, Damping $\alpha = 0.3$, Iterations $I = 8$, Variance floor $\epsilon = 10^{-4}$
\ENSURE Demodulated soft LLRs $\mathbf{L}_{tot}$
\STATE Let linear index $\mathbf{I}_{flat} = [1, 2, \dots, MN]^T$
\FOR{$p = 1$ \TO $P$}
    \STATE Pre-compute offline 1D LUTs: $\mathbf{idx}_{back}[:, p]$ and $\mathbf{idx}_{fwd}[:, p]$
    \STATE Pre-fetch shifted observation: $\mathbf{y}_{shift}[:, p] \leftarrow \mathbf{y}(\mathbf{idx}_{back}[:, p])$
\ENDFOR
\STATE Initialize local state: $\boldsymbol{\mu}_{x} \leftarrow \mathbf{0}, \, \boldsymbol{v}_{x} \leftarrow \mathbf{1}, \, \mathbf{L}_{edge} \leftarrow \mathbf{0}, \, \mathbf{L}_{tot} \leftarrow \mathbf{0}$
\STATE Initialize global observation: $\boldsymbol{\mu}_{y} \leftarrow \mathbf{0}, \, \boldsymbol{v}_{y} \leftarrow \left(\sigma^2_w + \sum_{p=1}^P |h_p|^2\right) \cdot \mathbf{1}$
\FOR{$iter = 1$ \TO $I$}
    \FOR{$p = 1$ \TO $P$ \quad (Loop Unrolled)}
        \STATE $\mathbf{i}_b \leftarrow \mathbf{idx}_{back}[:, p]$, $\quad \mathbf{i}_f \leftarrow \mathbf{idx}_{fwd}[:, p]$
        \STATE $\boldsymbol{\mu}_{int} \leftarrow \boldsymbol{\mu}_{y}(\mathbf{i}_b) - h_p \cdot \boldsymbol{\mu}_{x}[:, p]$
        \STATE $\boldsymbol{v}_{int} \leftarrow \max\left(\epsilon, \, \boldsymbol{v}_{y}(\mathbf{i}_b) - |h_p|^2 \cdot \boldsymbol{v}_{x}[:, p]\right)$
        \STATE $\mathbf{L}_{new} \leftarrow \left(4 \cdot \Re\left((\mathbf{y}_{shift}[:, p] - \boldsymbol{\mu}_{int}) \cdot h_p^*\right)\right) \oslash \boldsymbol{v}_{int}$
        \STATE $\mathbf{L}_{ext} \leftarrow \mathbf{L}_{tot} - \mathbf{L}_{edge}[:, p]$
        \STATE // \textit{Single-cycle hardware saturation clipping instead of $\tanh$}
        \STATE $\boldsymbol{\mu}_{x}[:, p] \leftarrow \max\left(-1.0f, \min\left(1.0f, \mathbf{L}_{ext} \cdot 0.5f\right)\right)$
        \STATE $\boldsymbol{v}_{x}[:, p] \leftarrow 1.0f - (\boldsymbol{\mu}_{x}[:, p])^2$
        \STATE // \textit{Momentum damping against loopy graph divergence}
        \STATE $\Delta \mathbf{L} \leftarrow \alpha \cdot \left(\mathbf{L}_{new} - \mathbf{L}_{edge}[:, p]\right)$
        \STATE $\mathbf{L}_{edge}[:, p] \leftarrow \mathbf{L}_{edge}[:, p] + \Delta \mathbf{L}$, $\quad \mathbf{L}_{tot} \leftarrow \mathbf{L}_{tot} + \Delta \mathbf{L}$
        \STATE Update global moments $\boldsymbol{\mu}_{y}(\mathbf{i}_f)$ and $\boldsymbol{v}_{y}(\mathbf{i}_f)$
    \ENDFOR
\ENDFOR
\RETURN $\mathbf{L}_{tot}$
\end{algorithmic}
\end{algorithm}

\subsection{1D Linearized Memory Refactoring}
Standard 2D OTFS detectors rely on nested loops and dynamic modulo arithmetic ($[l \pm \Delta l_p]_M$) to locate connected graph edges. On GPU SIMT architectures, such discontinuous indexing triggers massive uncoalesced global memory transactions and warp divergence. As outlined in Algorithm \ref{alg:mpa_detector}, we completely flatten the $M \times N$ grid into a contiguous 1D array of length $MN$. Multipath shifts are translated offline into static 1D forward and backward lookup tables (LUTs), eliminating branch divergence and forcing contiguous GPU memory coalescing.

\subsection{Momentum Damping and Hardware Saturation Clipping}
Updating variable node priors inside standard GA-MPA mandates evaluating transcendental hyperbolic tangent functions $\tanh(\cdot)$, consuming dozens of clock cycles per instruction inside CUDA Special Function Units (SFUs). Exploiting that uncoded BPSK symbols are bounded within $[-1, 1]$, we replace transcendental evaluations with a single-cycle hardware comparator instruction (\texttt{fmaxf}/\texttt{fminf}):
\begin{equation}
    \tanh\left(\frac{L_{ext}}{2}\right) \approx \max\left(-1.0f, \min\left(1.0f, \frac{L_{ext}}{2}\right)\right)
\end{equation}

Furthermore, belief propagation in deeply faded V2X channels frequently suffers from severe numerical oscillations. We embed a momentum damping mechanism ($\alpha = 0.3$) alongside a variance floor injection ($\epsilon = 10^{-4}$) to prevent hardware divide-by-zero exceptions, guaranteeing robust parallel execution.

\section{Experimental Validation and Profiling Analysis}

\subsection{Channel Emulator Verification and Empirical BER Analysis}
\begin{figure}[htbp]
    \centering
    \includegraphics[width=0.88\linewidth]{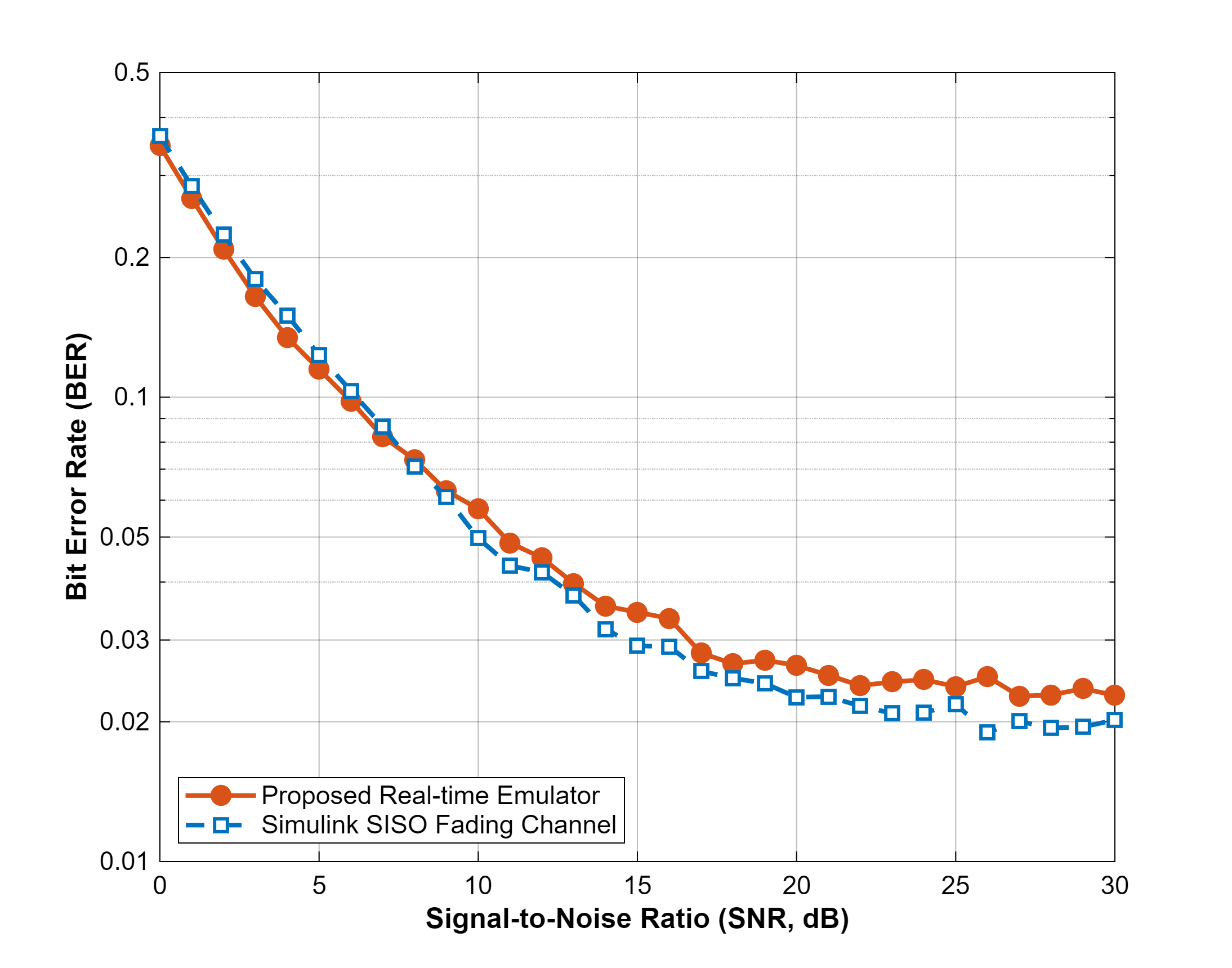}
    \caption{Empirical BER verification confirming numerical fidelity between our custom FP32 HIL channel emulator and the native FP64 module.}
    \label{fig:channel_verification}
\end{figure}

To verify experimental integrity, Fig. \ref{fig:channel_verification} benchmarks our custom FP32 real-time channel emulator against the native Simulink FP64 fading module under identical transceiver operations. Both engines demonstrate near-perfect statistical alignment across AWGN and EVA70 fading regimes (yielding $0.3481$ vs. $0.3650$ at $0\text{ dB}$, and converging to $0.0228$ vs. $0.0202$ at $30\text{ dB}$), confirming that FP32 quantization preserves orthogonal multipath phase fidelity.

\begin{figure}[htbp]
    \centering
    \includegraphics[width=0.92\linewidth]{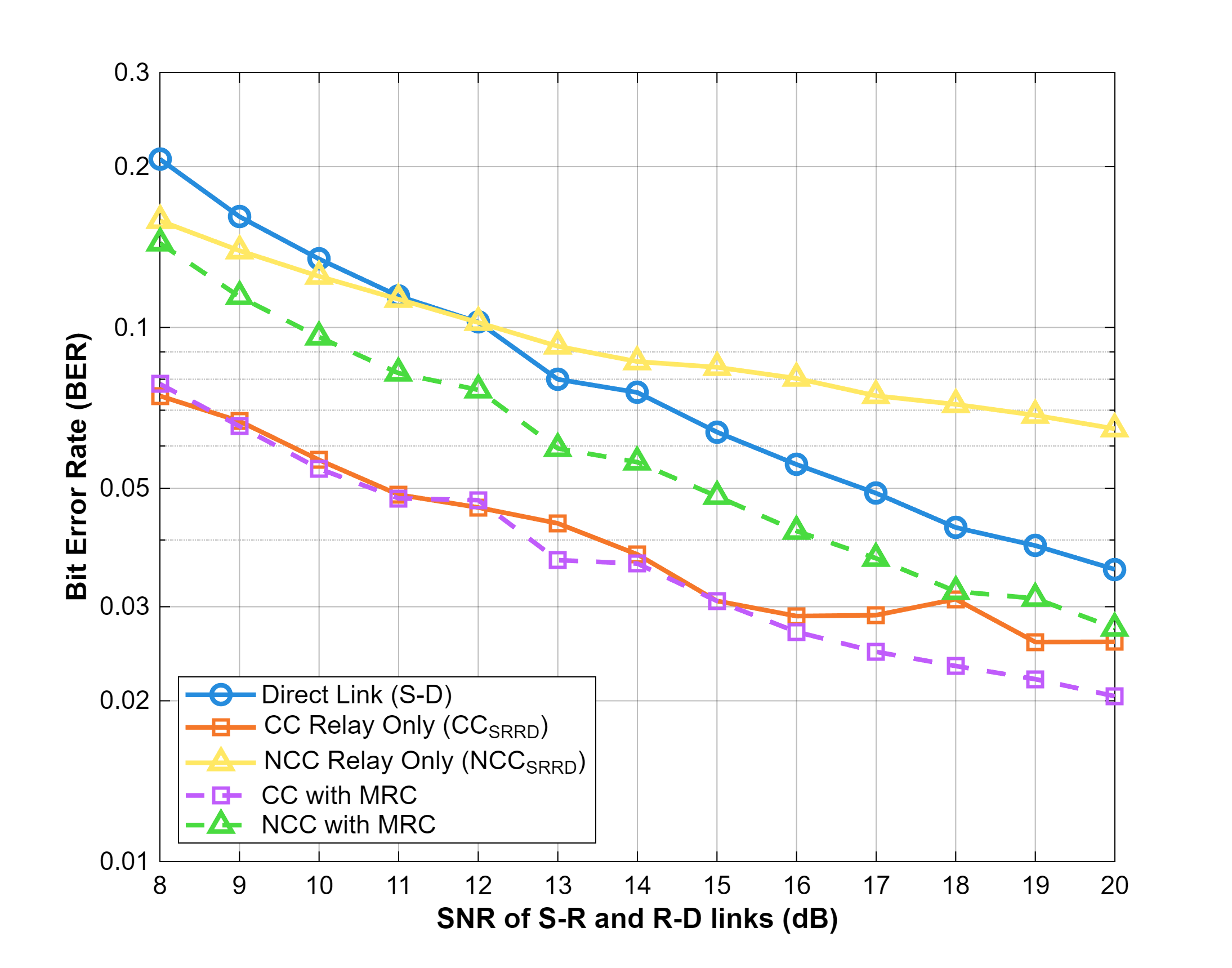}
    \centerline{\small (a) Theoretical Simulation: Perfect Relaying Assumption}
    \vspace{4mm}
    \includegraphics[width=0.92\linewidth]{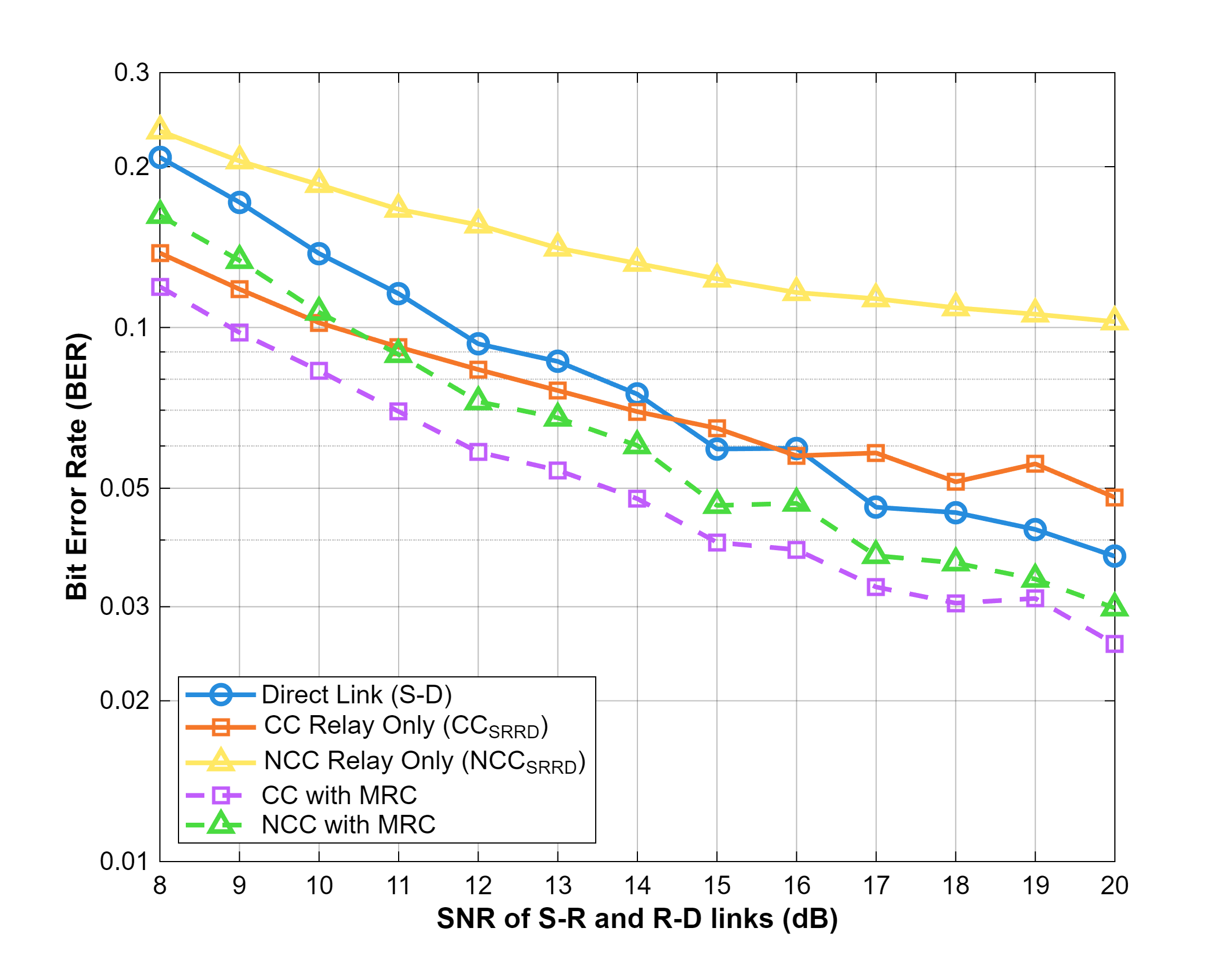}
    \centerline{\small (b) Theoretical Simulation: Imperfect Relaying Assumption}
    \caption{Monte Carlo BER simulation trajectories evaluating baseline GA-MPA decoding resilience across cooperative TDD topologies.}
    \label{fig:sim_ber}
\end{figure}

\begin{figure}[htbp]
    \centering
    \includegraphics[width=0.92\linewidth]{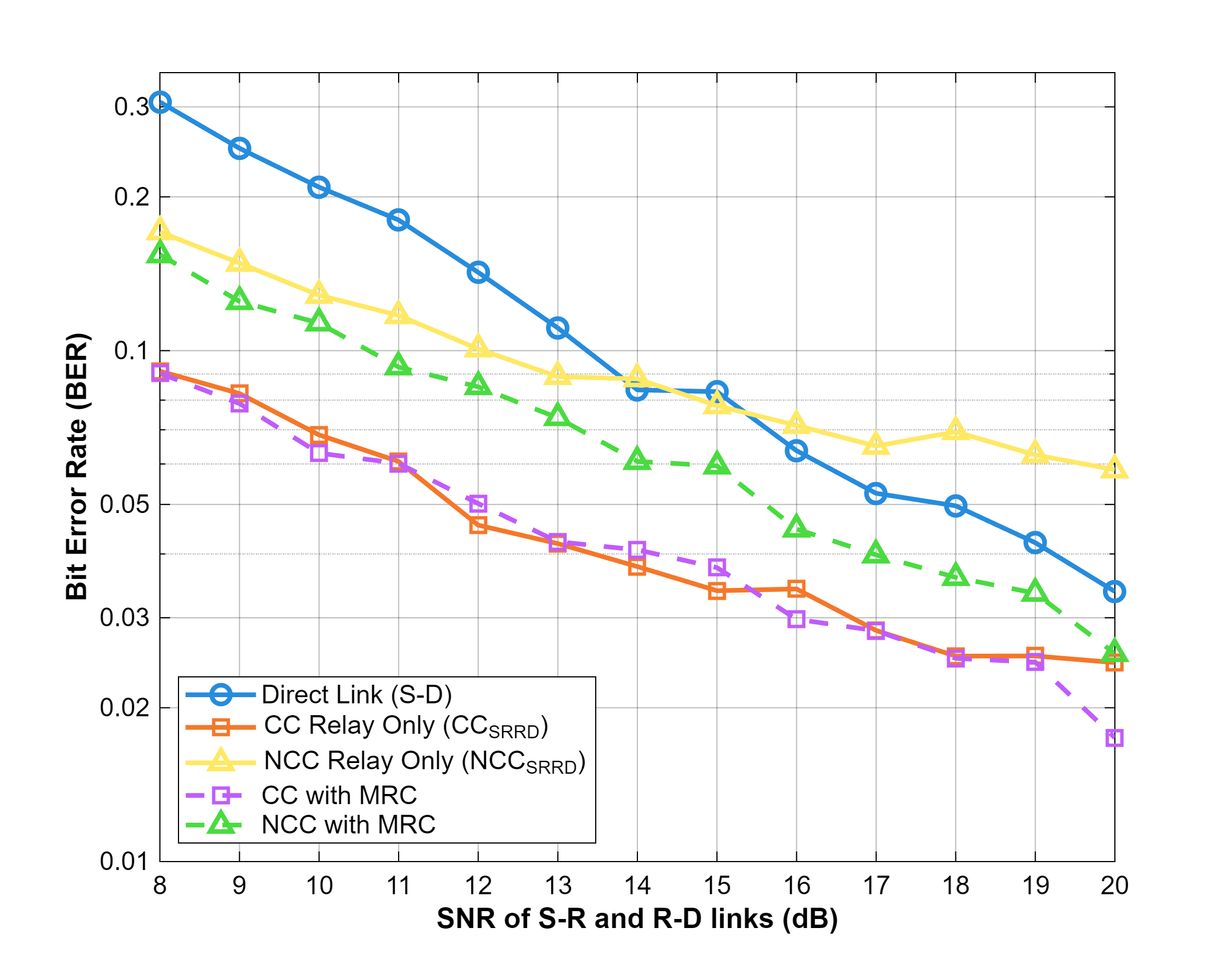}
    \centerline{\small (a) RF-Conducted SDR: Perfect Relaying Benchmark}
    \vspace{1mm}
    \includegraphics[width=0.92\linewidth]{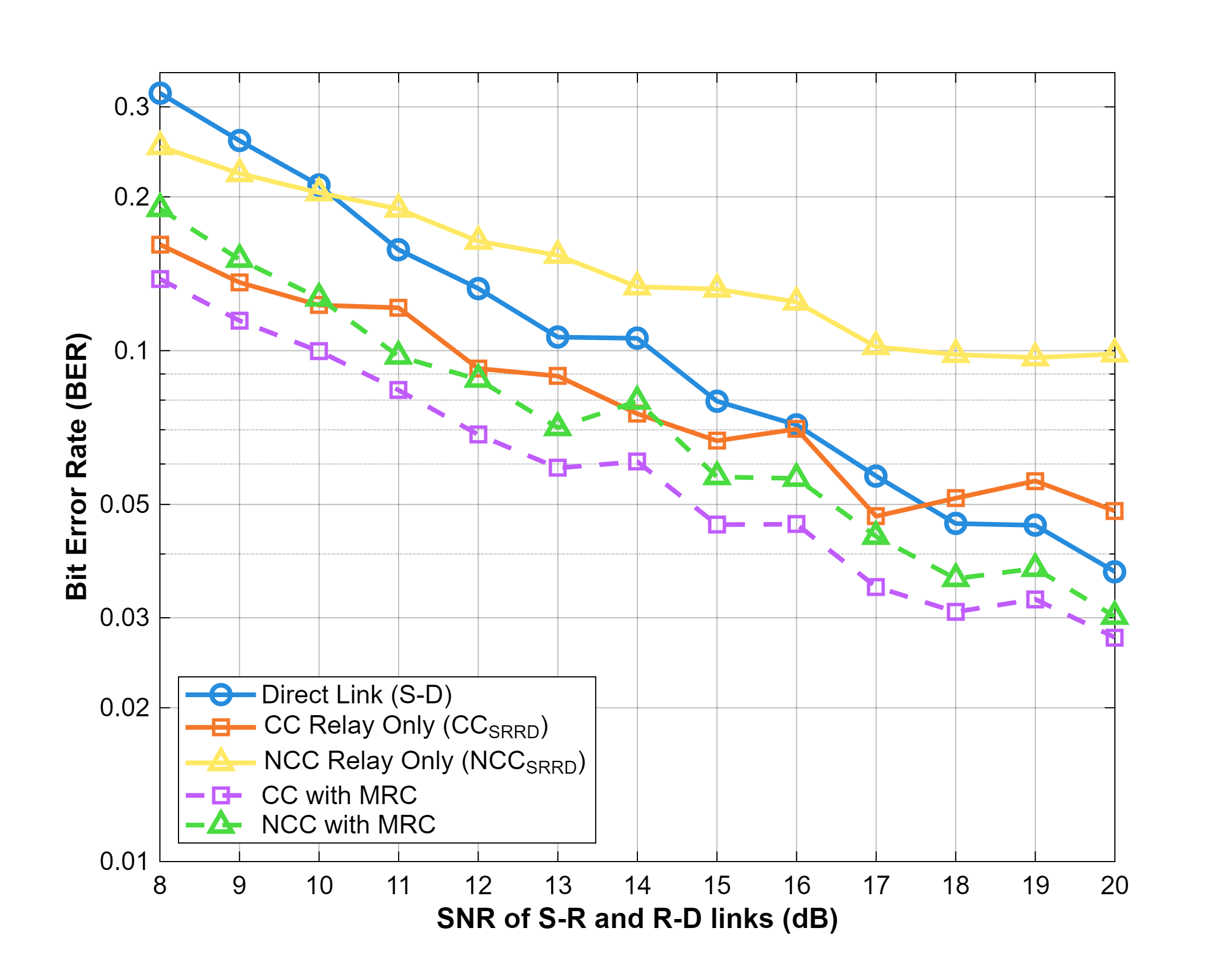}
    \centerline{\small (b) RF-Conducted SDR: Imperfect Relaying Benchmark}
    \caption{Empirical RF-conducted SDR BER validation over USRP B210 testbeds under HIL 3GPP EVA70 doubly dispersive fading.}
    \label{fig:sdr_ber}
\end{figure}

To evaluate end-to-end cooperative resilience, we analyze Bit Error Rate (BER) trajectories across Monte Carlo simulation (Fig. \ref{fig:sim_ber}) and physical RF-conducted HIL USRP B210 hardware transmissions (Fig. \ref{fig:sdr_ber}), contrasting \textit{Perfect Relaying} against \textit{Imperfect Relaying} (where $S \to R$ link errors naturally propagate through the emulated vehicular channel).

When deployed over physical SDR hardware under imperfect relaying (Fig. \ref{fig:sdr_ber}b), compound RF PA non-linearities and unmitigated relay error propagation cause standalone NCC reception ($\text{NCC}_{SRRD}$) to severely saturate at an empirical BER floor of $0.0984$, rendering it inferior even to the degraded direct link ($\text{SD}$, $0.0369$). However, via equivalent noise variance inflation, our proposed GA-MPA decoder dynamically down-weights corrupted relay LLRs during joint MRC ($\text{NCC}_{MRC}$). Consequently, joint $\text{NCC}_{MRC}$ successfully recovers the hardware error floor, pulling the empirical BER down to $0.0301$ at $20\text{ dB}$. This performance closely matches the ideal 4-slot cooperative benchmark ($0.0274$) with a negligible $0.0027$ margin, proving that OTFS-NCC improves spectral efficiency by 33\% while retaining full spatial diversity over real RF hardware.

\subsection{Heterogeneous RTF Profiling and Decoupling Analysis}
To quantify the intrinsic computational baseline of the transceiver pipeline, we execute a 120-second continuous signal transmission benchmark within the MATLAB/Simulink environment using the Simulink Profiler. As detailed in Table \ref{tab:profiling_breakdown}, executing standard 2D GA-MPA on a multi-core CPU results in severe processing congestion ($\text{RTF} = 4.37$). Crucially, brute-force migration of the naive 2D algorithm to the GPU only reduces the RTF to $1.60$, validating that brute-force compute scaling alone cannot resolve DD-domain memory bandwidth bottlenecks.

\begin{table}[htbp]
    \centering
    \caption{Simulink Model-Based RTF Profiling Breakdown ($120\text{ s}$ Simulated Signal Duration)}
    \resizebox{\columnwidth}{!}{%
    \begin{tabular}{l c c c c}
        \toprule
        \multirow{2}{*}{\textbf{Subsystem Module Name}} & \multicolumn{2}{c}{\textbf{2D Grid (Baseline)}} & \multicolumn{2}{c}{\textbf{1D Linearized (Alg.~\ref{alg:mpa_detector})}} \\
        \cmidrule(lr){2-3} \cmidrule(lr){4-5}
         & \textbf{Pure CPU} & \textbf{GPU Accel.} & \textbf{Pure CPU} & \textbf{GPU Accel.} \\
        \midrule
        EVA70 HIL Emulator [s] & 243.75 & 21.22 & 240.57 & 36.59 \\
        GA-MPA Detector [s] & 221.97 & 114.04 & 56.26 & 29.66 \\
        Other Baseband / Sync [s] & 58.39 & 56.19 & 55.46 & 40.97 \\
        \midrule
        \textbf{Total Execution Time [s]} & \textbf{524.11} & \textbf{191.46} & \textbf{352.30} & \textbf{107.22} \\
        \textbf{Real-Time Factor (RTF)} & \textbf{4.37} & \textbf{1.60} & \textbf{2.94} & \textbf{0.89} \\
        \bottomrule
    \end{tabular}%
    }
    \label{tab:profiling_breakdown}
\end{table}

By decoupling computational profiling, we isolate a hardware compute acceleration ratio of $1.95\times$ (2D) and $1.90\times$ (1D) provided by CUDA SFUs. Simultaneously, our 1D linearized memory mapping delivers a massive $3.95\times$ (CPU) and $3.85\times$ (GPU) structural speedup by eliminating warp divergence. Only when bridging heterogeneous GPU acceleration with 1D memory refactoring does the transceiver achieve true real-time performance, compressing execution to $107.22\text{ s}$ ($\text{RTF} = 0.89 < 1$).

\section{Conclusion}
This paper presented the first real-time SDR implementation of a Decode-and-Forward OTFS Network-Coded Cooperation system. By devising a hardware-algorithm decoupled acceleration architecture comprising 1D linearized memory mapping and transcendental function clipping, we compressed the equalization latency of a large $112 \times 64$ grid to a benchmarked RTF of $0.89$ in software simulation. RF-conducted HIL validation on USRP B210 testbeds confirmed that our framework effectively improves spectral efficiency by 33\% while suppressing relay error propagation under real EVA70 doubly dispersive fading.

\section*{Acknowledgements}
The authors acknowledge the use of generative AI tools (specifically Google Gemini and OpenAI ChatGPT) solely for English language polishing, code syntax check, and debugging assistance during the software implementation phase. All core theoretical formulations, protocol designs, system architectures, algorithm derivations, and empirical HIL validations were entirely conceived, executed, and verified by the authors, who take full and sole responsibility for the contents of this paper.

\bibliographystyle{IEEEtran}
\bibliography{references}

\end{document}